\documentclass[pra,preprint,aps]{revtex4}
\usepackage{graphics}

\begin{document}
\renewcommand{\textfraction}{0.2}
\renewcommand{\topfraction}{0.8}
\renewcommand{\bottomfraction}{0.8}
\renewcommand{\floatpagefraction}{0.3}

\title{High-precision evaluation of the Vibrational spectra of long-range
molecules}
\author{C. Tannous}
\affiliation{Laboratoire de Magnétisme de Bretagne, UPRES A CNRS 6135,
Université de Bretagne Occidentale, BP: 809 Brest CEDEX, 29285 FRANCE}
\author{J. Langlois}
\affiliation{Laboratoire des Collisions Electroniques et Atomiques,
Université de Bretagne Occidentale, BP: 809 Brest CEDEX, 29285 FRANCE}

\date{March 13, 2001}

\begin{abstract}
Vibrational spectra of long-range molecules are determined accurately
and to arbitrary accuracy with the Canonical Function Method. The
energy levels of the $0^-_g$ and $1_u$ electronic states of the
$^{23}{\rm Na}_2$ molecule are determined from the Ground state up to the
continuum limit. The method is validated by comparison with previous
results obtained by Stwalley et al. \cite {Stwalley 78} using the same
potential and  Trost et al. \cite {Trost 98} whose
 work is based on the Lennard-Jones potential adapted to long-range molecules.

\end{abstract}

\maketitle

\section{Introduction}
A new kind of high precision molecular spectroscopy is probing
long-range forces between constituent atoms of molecules. This
spectroscopy is based on using light to combine two colliding
cold-trapped atoms into a tenuous molecule. 

The burgeoning field of "Photoassociation Spectroscopy" is allowing
very  precise measurement of lifetimes of the first excited states of
Alkaline atoms and observation of retardation effects and long-range
forces. It provides a means of probing accurately the weak
interaction between these atoms \cite{Jones 96}.

The agreement between theory and experiment requires simultaneously a
highly accurate representation of the interaction potential as well as
a highly reliable method for the calculation of the corresponding
energy levels.

Since our aim is directed towards the latter problem, we make use of an
alternative method to evaluate the energy levels for the potential at 
hand instead of comparing to the experimental values in order to assess
the validity of our results.

The determination of the vibrational spectra of these very tenuous
molecules is extremely subtle specially for the highest levels which
play an important role in photoassociation spectroscopy. Thus a careful
control of accuracy is needed in order to diagonalise the Hamiltonian
without losing accuracy for all energies including those close to the
dissociation limit.

The magnitudes of potential energy, distance and mass values in these
kinds of molecules stand several orders of magnitude above or below what
is encountered in ordinary short-range molecules.

For instance, the typical intramolecular potential well depth at the
equilibrium distance of about 100 $a_0$ (Bohrs), is a fraction of a
cm$^{-1}$ while the reduced mass is several 10,000 electron masses. All
these extreme values require special numerical techniques in order to
avoid roundoffs, divergences, numerical instability and
ill-conditioning during processing.

The method we use in this work adapts well to this extreme situation
with the proviso of employing a series of isospectral scaling
transformations we explain below.

Since accuracy and its control are of paramount importance in this
work, the canonical function method (CFM) \cite {Kobeissi 82} is an
excellent candidate because it bypasses the calculation of the
eigenfunctions. This avoids losing accuracy associated with the
numerical calculation specially with rapidly oscillating wave functions
of highly excited states.

This method evaluates the full spectrum of the Hamiltonian, for any
potential, to any desired accuracy up to the continuum limit. It has
been tested succesfully in long-range and short-range potentials for
atomic and molecular states. It describes faithfully bound and free
states and is staightforward to code.

This work is organised as follows:  Section 2 is a description of the
CFM and highlights the details of the method. Section 3 presents the
results we obtain for the vibrational levels of the $^{23}{\rm Na}_2$
molecule $0^-_g$ and $1_u$ electronic states. Section 4 is an
additional validation of the method with the Lennard-Jones molecular
potential used in a similar situation. We conclude in section 5.

\section{The Canonical Function Method}
In this work we consider rotationless long-range diatomic molecules
only. The associated Radial Schr\"odinger equation (RSE) is given by:
\begin{equation}
(-\frac{\hbar^2}{2\mu}\frac{d^2}{dr^2}-V(r)+E)\psi(r)=0
\end{equation}
where $\mu$ is the reduced mass and $V(r)$ is the potential energy of the
interacting constituent atoms.

The diagonalisation of the RSE is described mathematically as a
singular boundary value problem. The prime advantage of the CFM is in
turning it into a regular initial value problem.

Primarily developed by Kobeissi and his coworkers \cite {Kobeissi 82}
the CFM, consists, essentially, in writing the general solution as a
function of the radial distance $r$ in terms of two basis functions
$\alpha(E;r)$ and $\beta(E;r)$ for some energy $E$.

Picking an arbitrary point $r_0$ at which a well defined set of initial
conditions are chosen \cite {Kobeissi 90}
i.e.: $\alpha(E;r_0)=1$ with $\alpha'(E;r_0)=0$ and $\beta(E;r_0)=0$
with $\beta'(E;r_0)=1$, the RSE is solved by progressing simultaneously towards the
origin and $\infty$. The effect of using different algorithms
for the numerical integration will be examined in the next section.

During the integration, the ratio of the functions is monitored until
saturation signaling the  stability of the eigenvalue spectrum \cite
{Kobeissi 91}.

One may define two associated energy functions:
\begin{equation}
l_{+}(E)=\lim_{r \rightarrow +\infty} -\frac{\alpha(E;r)}{\beta(E;r)};
\end{equation}
and:
\begin{equation}
l_{-}(E)= \lim_{r \rightarrow 0} -\frac{\alpha(E;r)}{\beta(E;r)}
\end{equation}
The eigenvalue function is defined in terms of:
\begin{equation}
F(E)=l_{+}(E)-l_{-}(E)
\end{equation}
The saturation of the $\alpha(E;r)/\beta(E;r)$ ratio as r
progresses towards 0 or $\infty$ yields a position independant
eigenvalue function $F(E)$. Its zeroes yield the spectrum of the RSE.

Generally, one avoids using the wavefunction but if one needs it, the
functions $\alpha(E;r)$ and $\beta(E;r)$ are used to determine the
wavefunction at any energy with the expression:
\begin{equation}
\Psi(E;r)=\Psi(E;r_0) \alpha(E;r) + \Psi'(E;r_0) \beta(E;r)
\end{equation}
where $\Psi(E;r_0)$ and $\Psi'(E;r_0)$ are the wavefunction and its
derivative at the start point $r_0$. The eigenfunctions are obtained
for $E=E_k$ where $E_k$ is any zero of $F(E)$.

\begin{figure}[htbp]
\begin{center}
\scalebox{0.5}{\includegraphics*{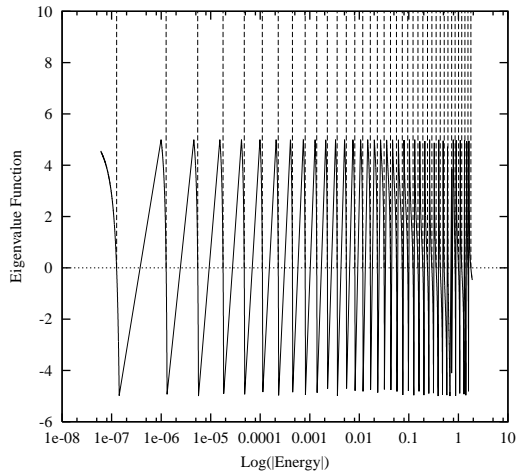}}
\end{center}
\caption{Behavior of the eigenvalue function $F(E)$ with energy on a
semi-log scale for the Vibrational spectra of the $0^-_g$ electronic
state of the $^{23}{\rm Na}_2$ molecule. The vertical lines indicate the
eigenvalue position. Energies are in cm$^{-1}$.} \label{fig1}
\end{figure}

The graph of the eigenvalue function has a typical $\tan(|E|)$ shape versus
the logarithm of the absolute value of the energy $E$ as displayed
in Fig.~\ \ref{fig1}.

\section{Vibrational states of the $^{23}{\rm Na}_2$ molecule}
We apply the CFM to the calculation of the vibrational energy levels of
a diatomic molecule where the interaction between the atoms is given by
the Movre and Pichler potential \cite {Movre 77}.

We start with the $0^-_g$ electronic state of the $^{23}{\rm Na}_2$
molecule. The corresponding potential is given by:
\begin{equation}
V(r)=\frac{1}{2}[(1-3X)+\sqrt{1-6X+81X^2}]
\end{equation}
where $X={C(0^-_g)}/{9r^3\Delta}$. $r$ is the internuclear distance
and the parameter $C(0^-_g)$ is such that:
\begin{equation}
\lim_{r \rightarrow +\infty} V(r) \rightarrow -\frac{C_3(0^-_g)}{r^3}
\end{equation}
Identification of the large $r$ limit yields the result:
$C_3(0^-_g)=C(0^-_g)/3$. We have used in the calculations below
$C_3(0^-_g)=6.390$ Hartrees.$a_0^3$ like \cite{Stwalley 78}. The
parameter $\Delta=1.56512.10^{-4}$ Rydbergs is the atomic spin-orbit
splitting. Given $C_3(0^-_g)$ and $\Delta$, the equilibrium
internuclear distance is $r_e=71.6 a_0$.

We scale all energies with a factor $E_0$ (usually cm$^{-1}$) with the
use of equation (1). Then we scale all distances with a typical length
$L_0$ transforming the RSE appropriately. This double transformation is
reflected generally in the potential coefficients preserving thus the
functional form of the potential. 

In order to gauge the accuracy of the spectra, we perform the
integration of the RSE with two different methods: A fixed step
Fourth-order Runge-Kutta method (RK4) and a Variable Step Controlled
Accuracy (VSCA) method \cite {Kobeissi 91}. 

The VSCA method is based on a series expansion of the potential and the
corresponding solution to an order such that a required tolerance
criterion is met. Ideally, the series coefficients are determined
analytically to any order, otherwise loss of accuracy occurs leading
quickly to numerical uncertainties as discussed later.

\begin{table}[htbp]
\begin{center}
\caption{Vibrational levels for the $0^-_g$ electronic state of the
$^{23}{\rm Na}_2$ molecule as obtained with a fixed step RK4 method,
Stwalley et al. \cite{Stwalley 78} results and the corresponding ratio.
Levels 34-40 were not found by the RK4 method due to the precision
limited to fourth order.} \label{tab1}
\begin{tabular}{ c c c c }
\hline* Index&  RK4 (cm$^{-1}$)&  Stwalley et al. (cm$^{-1}$)&
Ratio  \\ \hline
  1&    -1.7864563&   -1.7887&    1.00126\\
   2&   -1.5595812&   -1.5617&    1.00136\\
   3&   -1.3546211&   -1.3566&    1.00146\\
   4&   -1.1704091&   -1.1723&    1.00162\\
   5&   -1.0057168&   -1.0075&    1.00177\\
   6&   -0.8592746&   -0.86087&    1.00186\\
   7&   -0.7297888&   -0.73125&    1.00200\\
   8&   -0.6159592&   -0.61729&    1.00216\\
   9&   -0.5164938&   -0.51770&    1.00234\\
   10&  -0.4301231&   -0.43120&    1.00250\\
   11&  -0.3556114&   -0.35657&    1.00270\\
   12&  -0.2917683&   -0.29261&    1.00288\\
   13&  -0.2374567&   -0.23820&    1.00313\\
   14&  -0.1916004&   -0.19224&              1.00334\\
   15&  -0.1531898&   -0.15374&              1.00359\\
   16&  -0.1212858&        -0.12176&         1.00391\\
   17&  -9.5022481(-02)&    -9.5438(-02)&    1.00437\\
   18&  -7.3608067(-02)&    -7.3940(-02)&    1.00451\\
   19&  -5.6325397(-02)&    -5.6599(-02)&    1.00486\\
   20&  -4.2530440(-02)&    -4.2754(-02)&    1.00526\\
   21&  -3.1650256(-02)&    -3.1831(-02)&    1.00571\\
   22&  -2.3180032(-02)&    -2.3323(-02)&    1.00617\\
   23&  -1.6679434(-02)&    -1.6791(-02)&    1.00669\\
   24&  -1.1768423(-02)&    -1.1854(-02)&    1.00727\\
   25&  -8.1226859(-03)&    -8.1873(-03)&    1.00795\\
   26&  -5.4687973(-03)&    -5.5165(-03)&    1.00872\\
   27&  -3.5792655(-03)&    -3.6136(-03)&    1.00959\\
   28&  -2.2675456(-03)&    -2.2916(-03)&    1.01061\\
   29&  -1.3831324(-03)&    -1.3995(-03)&    1.01183\\
   30&  -8.0680818(-04)&    -8.1747(-04)&    1.01321\\
   31&  -4.4611287(-04)&    -4.5276(-04)&    1.01490\\
   32&  -2.3077899(-04)&    -2.3503(-04)&    1.01842\\
   33&  -9.4777816(-05)&    -1.1252(-04)&    1.18720\\
   34& 	                   &-4.8564(-05)&               \\
   35& 		           &-1.8262(-05)&               \\
   36&	 	           &-5.6648(-06)&               \\
   37& 		           &-1.3175(-06)&             \\
   38&	 	           &-1.9247(-07)&               \\
   39& 		           &-1.1215(-08)&             \\
   40&	 	           &-4.1916(-11)&              \\

\hline*
\end{tabular}
\end{center}
\end{table}

Table \ref{tab1} shows the results we obtain with the RK4 method.
The limitation of the RK4 method to fourth order hampers the finding of
levels beyond the 33rd (see Table 1). In order to find the higher
levels we have to select an algorithm that enables us to tune the
accuracy well beyond the fourth order.

Pushing the accuracy within the framework of a fixed step method has
the effect of reducing substantially the integration step. In order to
avoid this problem, we use a variable step that adjusts itself to the
desired accuracy, the VSCA method.

This method is powerful and flexible enough to find all the desired energy
levels and allows us to find one additional level that was not detected before. 
It should be noted that the last three levels given by Stwalley at al.
\cite{Stwalley 78} were extrapolated and not calculated. The agreement
between our calculated levels and those of Stwalley et al.
\cite{Stwalley 78} is quite good. We believe that the small
discrepancy, increasing as we progress towards the dissociation limit,
is due to a loss of accuracy associated with traditional methods in
sharp contrast with the CFM.

\begin{table}[htbp]
\begin{center}
\caption{Vibrational levels for the $0^-_g$ electronic state of the
$^{23}{\rm Na}_2$ molecule as obtained with Stwalley et al. results, the
variable step controlled accuracy method (VSCA) method and the
corresponding ratio. Levels 38, 39 and 40 of Stwalley et al.
\cite{Stwalley 78} are extrapolated with LeRoy and Bernstein \cite{LeRoy
70} semi-classical formulae.} \label{tab2}
\begin{tabular}{ c c c c }
\hline* Index& Stwalley et al. (cm$^{-1}$) & VSCA (cm$^{-1}$) & Ratio \\
\hline
  1&   -1.7887&   -1.7864488&             1.00126\\
   2&  -1.5617&   -1.5595638&             1.00137\\
   3&  -1.3566&   -1.3546072&             1.00147\\
   4&  -1.1723&   -1.1703990&             1.00162\\
   5&  -1.0075&   -1.0057071&              1.00178\\
   6&  -0.86087&   -0.8592631&             1.00187\\
   7&  -0.73125&   -0.7297908&             1.00200\\
   8&  -0.61729&   -0.6159534&             1.00202\\
   9&  -0.51770&   -0.5164882&             1.00235\\
  10&  -0.43120&   -0.4301217&             1.00251\\
  11&  -0.35657&   -0.3556148&             1.00269\\
  12&  -0.29261&   -0.2917693&             1.00288\\
  13&  -0.23820&   -0.2374560&             1.00313\\
  14&  -0.19224&   -0.1916002&             1.00334\\
  15&  -0.15374&   -0.1531893&             1.00359\\
  16&  -0.12176&         -0.1212854&       1.00391\\
  17&  -9.5438(-02)&     -9.5022588(-02)&   1.00437\\
  18&  -7.3940(-02)&     -7.3608452(-02)&   1.00450\\
  19&  -5.6599(-02)&     -5.6325744(-02)&   1.00485\\
  20&  -4.2754(-02)&     -4.2530867(-02)&   1.00525\\
  21&  -3.1831(-02)&     -3.1650591(-02)&   1.00570\\
  22&  -2.3323(-02)&     -2.3180420(-02)&   1.00615\\
  23&  -1.6791(-02)&     -1.6679756(-02)&   1.00667\\
  24&  -1.1854(-02)&     -1.1768655(-02)&   1.00725\\
  25&  -8.1873(-03)&     -8.1228816(-03)&   1.00793\\
  26&  -5.5165(-03)&     -5.4689541(-03)&   1.00869\\
  27&  -3.6136(-03)&     -3.5793742(-03)&   1.00956\\
  28&  -2.2916(-03)&     -2.2676168(-03)&   1.01058\\
  29&  -1.3995(-03)&     -1.3831806(-03)&   1.01180\\
  30&  -8.1747(-04)&     -8.0683859(-04)&   1.01318\\
  31&  -4.5276(-04)&     -4.4613249(-04)&  1.01486\\
  32&  -2.3503(-04)&     -2.3110168(-04)&   1.01700\\
  33&  -1.1252(-04)&     -1.1035443(-04)&   1.01962\\
  34&  -4.8564(-05)&     -4.7468345(-05)&   1.02308\\
  35&  -1.8262(-05)&     -1.7767388(-05)&   1.02784\\
  36&  -5.6648(-06)&     -5.4747950(-06)&   1.03471\\
  37&  -1.3175(-06)&     -1.2597092(-06)&   1.04588\\
  38&  -1.9247(-07)&     -1.2716754(-07)&   1.51352\\
  39&  -1.1215(-08)&                                     \\
  40&  -4.1916(-11)&                                      \\
\hline*
\end{tabular}
\end{center}
\end{table}

The estimation of accuracy of the results hinges basically
on two operations, integration and determination of the zeroes
of the eigenvalue function $F(E)$. The superiority of the VSCA
method is observed in the determination of the upper levels that are 
not detected by the RK4 method (see Tables 1 and 2).
In addition, it is observed in the behavior of the eigenvalue ratio
versus the index. While in both cases (RK4 and VSCA) the ratio increases steadily as
the index increases because we are probing higher excited states, in the 
RK4 case it rather blows up as dissociation is approached. We use typically
series expansion to order 12 in VSCA with a tolerance of $10^{-8}$. 
In the root search of $F(E)$, the tolerance required for a zero
to be considered as an eigenvalue is $10^{-15}$.
This does not imply that we disagree as strongly as $0.13\%$, for instance,
with the Ground state value (see Tables 1 and 2) found by Stwalley et al.
\cite {Stwalley 78} for the simple reason, we use a splitting energy 
$\Delta=1.56512.10^{-4}$ Rydbergs  corresponding to an equilibrium
internuclear distance $r_e=71.6 a_0$. 
Stwalley et al. do not provide explicitly the value of $\Delta$ they use, 
and more recently Jones et al. \cite{Jones 96} provide a value that is slightly
different.\\

\begin{figure}[htbp]
\begin{center}
\scalebox{0.5}{\includegraphics*{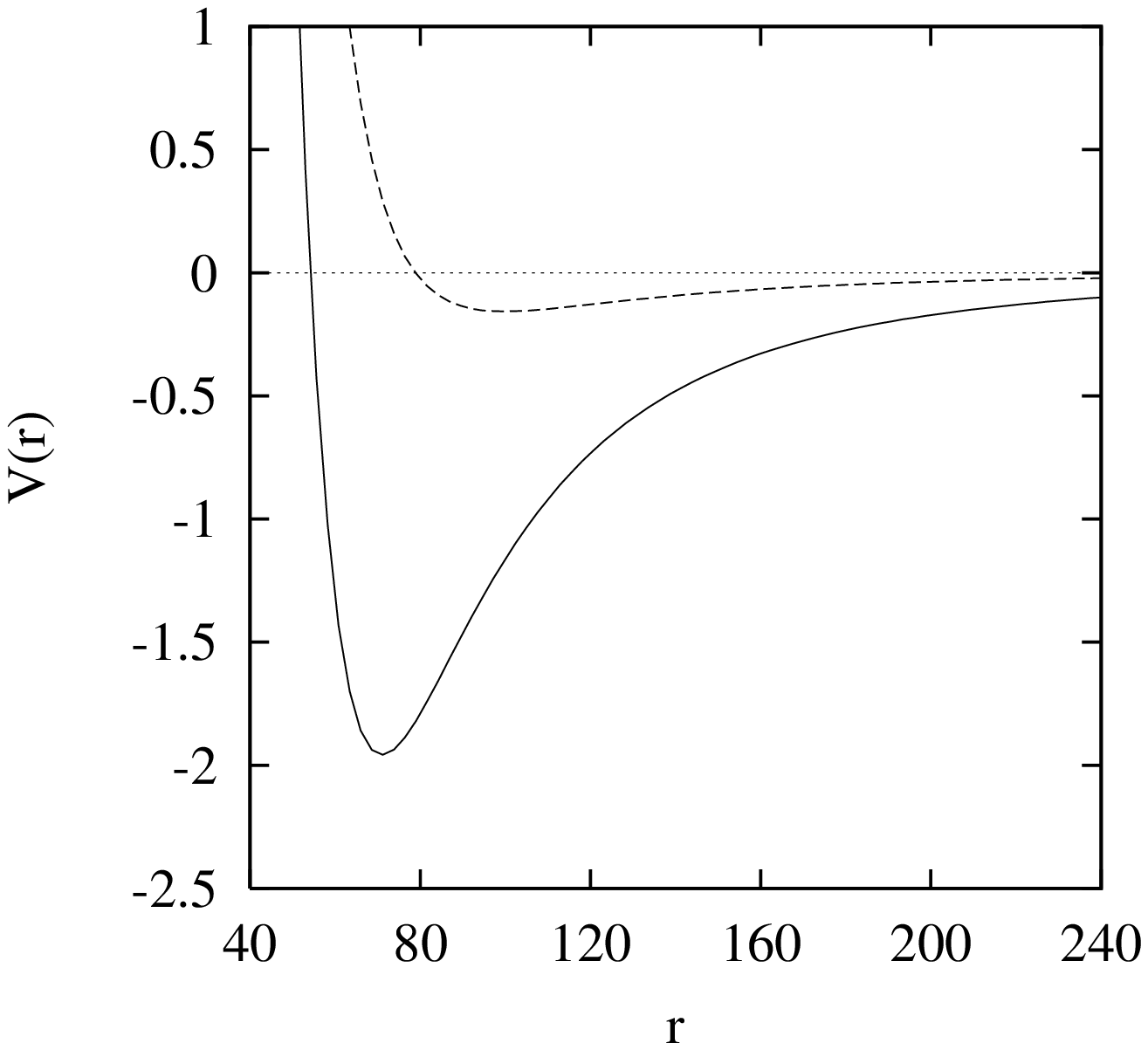}}
\end{center}
\caption{Potential energy in cm$^{-1}$ for the $0^-_g$ and $1_u$
electronic states of the $^{23}{\rm Na}_2$ molecule. The radial distance r
is in $a_0$ units.} \label{fig2}
\end{figure}

We treat next the $1_u$ electronic state of the $^{23}{\rm Na}_2$ molecule.
This state is higher that the $0^-_g$ and the number of vibrational
levels is smaller because the potential is shallower as displayed in
Fig.~\ \ref{fig2}.

The potential associated with the $1_u$ electronic state of the
$^{23}{\rm Na}_2$ molecule is a lot more involved. Its VSCA implementation
is particularly difficult because of the complex functional of the
potential as we explain below. Analytically, the VSCA algorithm
requires performing a Taylor series expansion to any order around an
arbitrary point \cite{Kobeissi 90}.

This is still, numerically, an open problem for arbitrary functions and
the use of LISP based symbolic manipulation techniques produces quickly
cumbersome expressions. Special methods based on analytical fitting
expressions are needed in order to turn the series coefficients into a
more manageable form.

The first step is to determine the $1_u$ electronic state of the
$^{23}{\rm Na}_2$ molecule by solving the Movre et al. \cite {Movre 77}
secular equation such that:
\begin{equation}
V(r)=\Delta[-2\sqrt{Q}cos(\frac{\theta-2\pi}{3})-\frac{a}{3}-1]
\end{equation}
where a=-2-6X and $X={C(1_u)}/{9r^3\Delta}$. In addition,
$\theta=cos^{-1}( \frac{1+270X^3}{\sqrt{(1+63X^2)^3}})$, and
$Q=\frac{1+63X^2}{9}$.

The parameter $C(1_u)$ is such that:
\begin{equation}
\lim_{r \rightarrow +\infty} V(r) \rightarrow -\frac{C_3(1_u)}{r^3}
\end{equation}

Identification of the large $r$ limit yields to the result:
$C_3(1_u)=C(1_u)(\sqrt{7}-2)/9$.
 We have used in the calculations below  $C_3(1_u)$=1.383 Hartrees.$a_0^3$ like
Stwalley et al. The parameter $\Delta=1.56512.10^{-4}$ Rydbergs is the
same as for the $0^-_g$ state.

Table \ref{tab3} displays the results we obtain with the RK4
method that cannot find more than 11 levels due to accuracy limitations.

\begin{table}[htbp]
\begin{center}
\caption{Vibrational levels for the $1_u$ electronic state of the
$^{23}{\rm Na}_2$ molecule as obtained with the RK4 method, Stwalley
et al. results and the corresponding ratio. RK4 found only 11 levels
and levels 15 and 16 of Stwalley et al. are found by extrapolation.}
\label{tab3}
\begin{tabular}{ c c c c }
\hline* Index&  RK4 (cm$^{-1}$) &  Stwalley et al.(cm$^{-1}$) &
Ratio \\ \hline
   1&    0.1319536     &    0.13212&        1.00126\\
   2&    9.0057392(-02)&    9.0192(-02)&    1.00149\\
   3&    5.9472159(-02)&    5.9574(-02)&    1.00171\\
   4&    3.7821597(-02)&    3.7896(-02)&    1.00197\\
   5&    2.3027828(-02)&    2.3080(-02)&    1.00227\\
   6&    1.3324091(-02)&    1.3359(-02)&    1.00262\\
   7&    7.2562182(-03)&    7.2787(-03)&    1.00310\\
   8&    3.6714971(-03)&    3.6849(-03)&    1.00365\\
   9&    1.6950002(-03)&    1.7024(-03)&    1.00437\\
   10&   6.9533077(-04)&    6.9904(-04)&    1.00533\\
   11&   2.4102039(-04)&    2.4492(-04)&    1.01618\\
   12&                 &    6.8430(-05)&           \\
   13&                 &    1.3446(-05)&           \\
   14&                 &    1.4122(-06)&           \\
   15&                 &    3.8739(-08) &         \\
   16&                 &    1.2735(-12) &            \\

\hline*
\end{tabular}
\end{center}
\end{table}

The next results for the $1_u$ electronic state of the $^{23}{\rm Na}_2$
molecule are obtained with the VSCA method as shown in  table \ref{tab4}. 
We find an additional 15-th level in contrast to  Stwalley et al. who found
fourteen and extrapolated the last two levels.

\begin{table}[htbp]
\begin{center}
\caption{Vibrational levels for the $1_u$ electronic state of the
$^{23}{\rm Na}_2$ molecule as obtained by Stwalley et al., the variable
step controlled accuracy method (VSCA) and the corresponding ratio. A
new 15th level is obtained with the VSCA method.} \label{tab4}
\begin{tabular}{ c c c c }
\hline* Index& Stwalley et al. (cm$^{-1}$) & VSCA (cm$^{-1}$) & Ratio \\
\hline
  1&     0.13212&        0.13244150&         1.00243\\
   2&    9.0192(-02)&    9.07598688(-02)&    1.00630\\
   3&    5.9574(-02)&    6.01645742(-02)&    1.00991\\
   4&    3.7896(-02)&    3.83982868(-02)&    1.01325\\
   5&    2.3080(-02)&    2.34584275(-02)&    1.01640\\
   6&    1.3359(-02)&    1.36187753(-02)&    1.01944\\
   7&    7.2787(-03)&    7.44243743(-03)&    1.02250\\
   8&    3.6849(-03)&    3.77999552(-03)&    1.02581\\
   9&    1.7024(-03)&    1.75281307(-03)&    1.02961\\
   10&   6.9904(-04)&    7.23013793(-04)&    1.03430\\
   11&   2.4492(-04)&    2.54856605(-04)&    1.04057\\
   12&   6.8430(-05)&    7.18249908(-05)&    1.04961\\
   13&   1.3446(-05)&    1.43120199(-05)&    1.06441\\
   14&   1.4122(-06)&    1.53931042(-06)&    1.09001\\
   15&   3.8739(-08)&    4.66022073(-09)&            \\
   16&   1.2735(-12)&                                \\
\hline*
\end{tabular}
\end{center}
\end{table}

The corresponding graph of the eigenvalue function is displayed in
Fig.~\ \ref{fig3} below.\\

\begin{figure}[htbp]
\begin{center}
\scalebox{0.5}{\includegraphics*{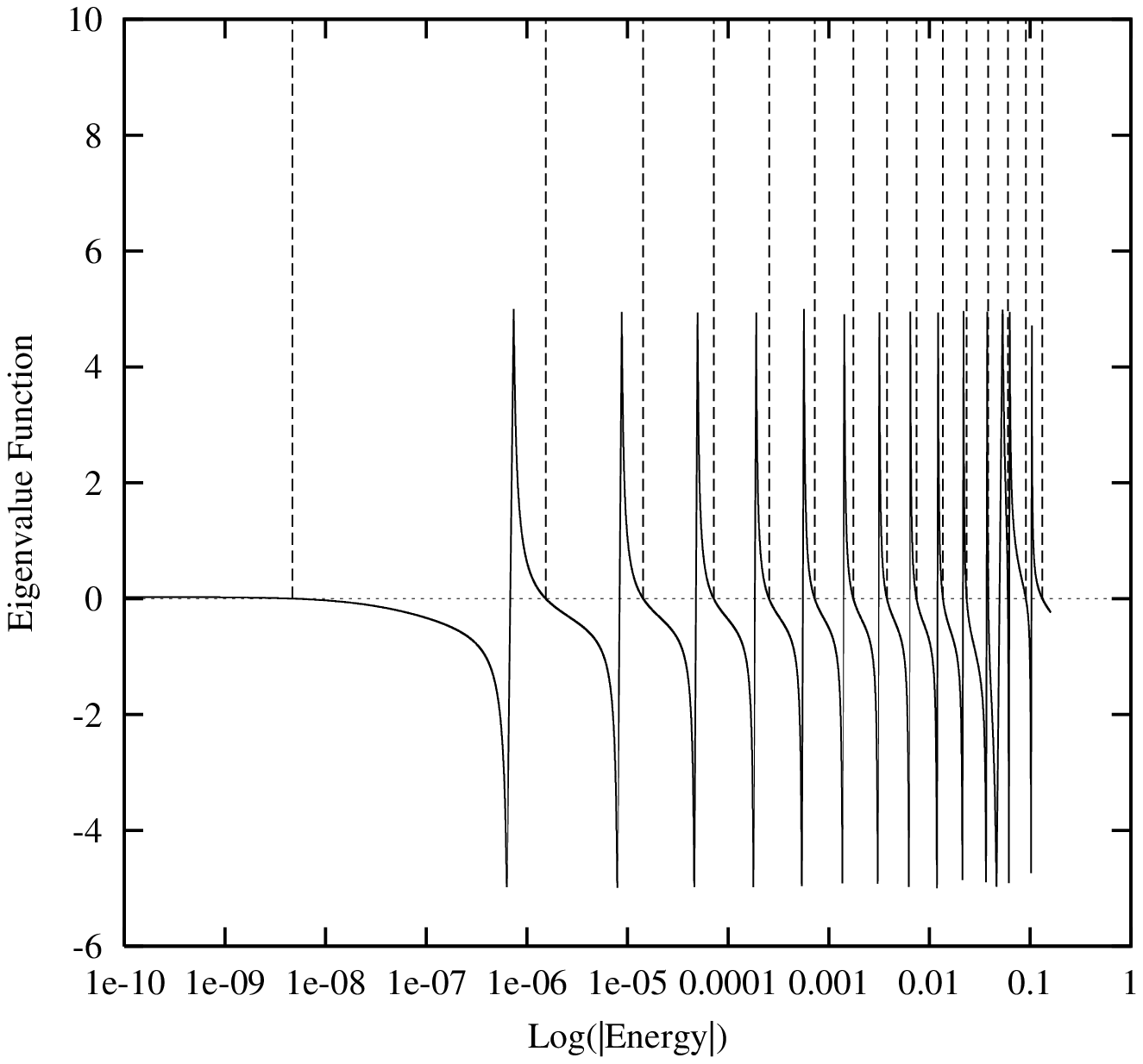}}
\end{center}
\caption{Behavior of the eigenvalue function $F(E)$ with energy on a
semi-log scale for the $1_u$ electronic state of the $^{23}{\rm Na}_2$
molecule. The vertical lines indicate the eigenvalue position. Energies
are in cm$^{-1}$.} \label{fig3}
\end{figure}

\section{Lennard-Jones molecules}
We apply our methodology to the Lennard-Jones case. Our results are
compared to the results obtained by Trost et al. \cite{Trost 98}. We
start with the levels obtained with the RK4 method. The energy unit is
$\epsilon$ the depth of the potential well of the Asymmetric
Lennard-Jones potential (ALJ):

\begin{equation}
V(r)=C_1 ({\frac{1}{r}})^\beta - C_2 ({\frac{1}{r}})^\alpha
\end{equation}

\begin{figure}[htbp]
\begin{center}
\scalebox{0.5}{\includegraphics*{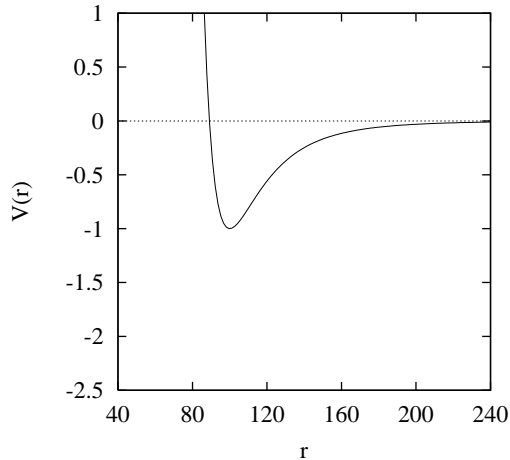}}
\end{center}
\caption{Potential energy of the Trost et al. \cite{Trost 98} model.
The energy is in $\epsilon$ units where $\epsilon$ is the potential
well depth and the radial distance is in $\frac{a_0}{\sqrt{B}}$ where
$B$ is a scaled mass.} \label{fig4}
\end{figure}

The Asymmetric-Lennard-Jones (ALJ) depends on $C_1$ and $C_2$ yielding
an equilibrium distance at $r=r_{min}$  and a potential depth of
$-\epsilon$. Trost et al. \cite{Trost 98} use the general
parameterisation:
\begin{equation}
C_1 = \frac{\epsilon}{(\beta-\alpha)}\alpha r_{min}^\beta
\hspace{0.1cm}, \hspace{0.1cm} C_2=\frac{\epsilon}{(\beta-\alpha)}\beta
r_{min}^\alpha
\end{equation}
It is scaled in such a way that the energy is expressed in units of the
potential well depth $\epsilon$. When $\alpha=6$, $\beta=12$ we obtain
$r_{min}=\sqrt[6]{\frac{2C_{1}}{C_2}}$ and $\epsilon=
\frac{C_2^2}{4C_{1}}$. and the radial distance is in
$\frac{a_0}{\sqrt{B}}$ where $B$ is a reduced scaled mass given by $B=2
\mu \epsilon {r_{min}}^2$. Numerically Trost et al. use $B=10^4$ which
is the order of magnitude encountered in long-range molecules within
the framework of their system of units.
The potential energy in these units is displayed in Fig.~\ \ref{fig4}.
The RK4 methods yields the results displayed in Table \ref{tab5}.

\begin{table}[htbp]
\caption{Quantum levels of a Lennard-Jones molecule in $\epsilon$
units, the depth of the potential well as obtained by the RK4 method,
Trost et al. and the corresponding ratio. Only the last 24th level was
missed by the RK4 method.} \label{tab5}
\begin{center}
\begin{tabular}{ c c c c }
\hline* Index& RK4 &  Trost et al.& Ratio \\ \hline
   1&   -0.9410450&      -0.9410460&    1.000001\\
   2&   -0.8299980&      -0.8300020&    1.000005\\
   3&   -0.7276400&      -0.7276457&    1.000008\\
   4&   -0.6336860&      -0.6336930&    1.000011\\
   5&   -0.5478430&      -0.5478520&    1.000017\\
   6&   -0.4698130&      -0.4698229&    1.000021\\
   7&   -0.3992870&      -0.3992968&    1.000025\\
   8&   -0.3359470&      -0.3359561&    1.000027\\
   9&   -0.2794670&      -0.2794734&           1.000023\\
  10&   -0.2295070&      -0.2295117&           1.000021\\
  11&   -0.1857220&      -0.1857237&           1.000009\\
  12&   -0.1477510&      -0.1477514&           1.000003\\
  13&   -0.1152270&      -0.1152259&           0.999990\\
  14&   -8.776970(-02)&  -8.7766914(-02)&      0.999968\\
  15&   -6.498640(-02)&  -6.4982730(-02)&      0.999944\\
  16&   -4.647400(-02)&  -4.6469911(-02)&      0.999912\\
  17&   -3.181750(-02)&  -3.1813309(-02)&      0.999868\\
  18&   -2.059000(-02)&  -2.0586161(-02)&      0.999814\\
  19&   -1.235370(-02)&  -1.2350373(-02)&      0.999731\\
  20&   -6.659580(-03)&  -6.6570240(-03)&      0.999616\\
  21&   -3.048890(-03)&  -3.0471360(-03)&      0.999425\\
  22&   -1.053690(-03)&  -1.0527480(-03)&      0.999106\\
  23&   -1.645210(-04)&  -1.9834000(-04)&      1.205560\\
  24&                 &  -2.6970000(-06) &              \\

\hline*
\end{tabular}
\end{center}
\end{table}

\begin{table}[htbp]
\begin{center}
\caption{Quantum levels of a Lennard-Jones molecule in $\epsilon$
units, the depth of the potential well as obtained by the VSCA method,
Trost et al. and the corresponding ratio.} \label{tab6}
\begin{tabular}{ c c c c }
\hline* Index& VSCA &  Trost et al. & Ratio \\ \hline
 1&    -0.9410443&        -0.9410460&        1.000002\\
   2&  -0.8299963&        -0.8300020&        1.000007\\
   3&  -0.7276415&        -0.7276457&        1.000006\\
   4&  -0.6336915&        -0.6336930&        1.000002\\
   5&  -0.5478480&        -0.5478520&        1.000007\\
   6&  -0.4698206&        -0.4698229&        1.000005\\
   7&  -0.3992947&        -0.3992968&        1.000005\\
   8&  -0.3359533&        -0.3359561&        1.000008\\
   9&  -0.2794718&        -0.2794734&        1.000005\\
  10&  -0.2295109&        -0.2295117&        1.000003\\
  11&  -0.1857222&        -0.1857237&        1.000008\\
  12&  -0.1477498&        -0.1477514&        1.000010\\
  13&  -0.1152247&        -0.1152259&        1.000010\\
  14&  -8.7766358(-02)&   -8.7766914(-02)&   1.000006\\
  15&  -6.4982534(-02)&   -6.4982730(-02)&   1.000003\\
  16&  -4.6469838(-02)&   -4.6469911(-02)&   1.000002\\
  17&  -3.1813146(-02)&   -3.1813309(-02)&   1.000005\\
  18&  -2.0585953(-02)&   -2.0586161(-02)&   1.000010\\
  19&  -1.2350173(-02)&   -1.2350373(-02)&   1.000016\\
  20&  -6.6568735(-03)&   -6.6570240(-03)&   1.000023\\
  21&  -3.0470500(-03)&   -3.0471360(-03)&   1.000028\\
  22&  -1.0526883(-03)&   -1.0527480(-03)&   1.000057\\
  23&  -1.9832170(-04)&   -1.9834000(-04)&   1.000092\\
  24&  -2.6957891(-06)&   -2.6970000(-06)&   1.000449\\
\hline*
\end{tabular}
\end{center}
\end{table}

\begin{figure}[htbp]
\begin{center}
\scalebox{0.5}{\includegraphics*{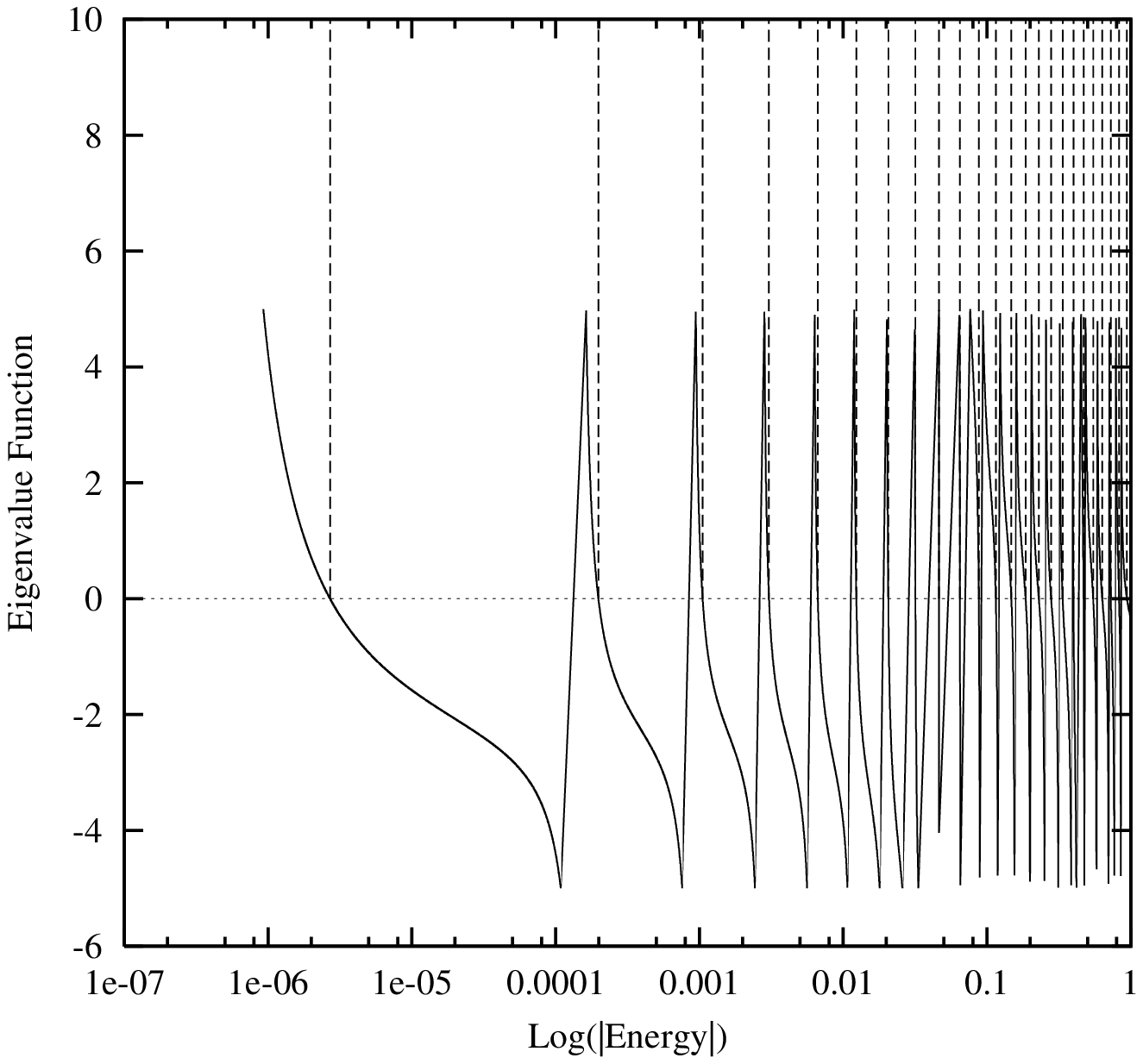}}
\end{center}
\caption{ Behavior of the eigenvalue function $F(E)$ with energy on a
semi-log scale for the Trost et al. \cite{Trost 98} Lennard-Jones
molecule. The vertical lines indicate the eigenvalue position. Energies
are in potential depth $\epsilon$ units.} \label{fig5}
\end{figure}

The accuracy limitation of the RK4 results in losing the uppermost
level 24. Thus we move on to the results obtained with the superior
VSCA method.
All levels are obtained with the VSCA and the agreement with Trost et
al. results is perfect as witnessed by the ratio values of Table 6.
The eigenvalue function obtained with the VSCA method as a function of
energy is displayed in Fig.~\ \ref{fig5}.

\section{Conclusion}
The CFM is a very powerful method that enables one to find the
Vibrational spectra of tenuous molecules where energies and distances
are so remote from the ordinary short-range molecules case that special
techniques should be developed in order to avoid numerical
instabilities and uncertainties.

The VSCA integration method used gives the right number of all the
levels and the variation of the eigenvalue function $F(E)$ definitely
determines the total number of levels. Generally it requires performing
analytically Taylor series expansion to any order of an arbitrary
potential function that might require the combination of numerical,
symbolic manipulation and functional fitting techniques. Despite these
difficulties, the results of the VSCA are rewarding.

The use of the RK4 and the VSCA methods jointly paves the way to a
precise comparative evaluation of the accuracy of the spectra obtained.
In practice, the RK4 method can be used in optimisation problems
whereas the VSCA is more adapted to the direct evaluation of the
spectra.

In this work, we did not consider molecular rotation, nevertheless the
CFM is adapted to solve accrately Ro-Vibrational problems as well as
any RSE diagonalisation problem.

The methodology we describe in this work enables one to tackle
precisely weakly bound states that are of great importance in
low-energy scattering of atoms and molecules and generally in
Bose-Einstein condensation problems \cite {Trost 98}.\\

{\bf Acknowledgements}: Part of this work was performed on the IDRIS
machines at the CNRS (Orsay).\\

\end{document}